\documentclass[]{spie}  

\usepackage{amsmath,amsfonts,amssymb}
\usepackage{graphicx}
\usepackage[colorlinks=true, allcolors=blue]{hyperref}

\title{Various Wavefront Sensing and Control Developments on the Santa Cruz Extreme AO Laboratory (SEAL) Testbed}

\author[a,b]{Benjamin L. Gerard}
\author[b]{Javier Perez-Soto}
\author[c]{Vincent Chambouleyron}
\author[a,b]{Maaike A.M. van Kooten}
\author[a]{Daren Dillon}
\author[d,e]{Sylvain Cetre}
\author[a,b]{Rebecca Jensen-Clem}
\author[f]{Qiang Fu}
\author[f]{Hadi Amata}
\author[f]{Wolfgang Heidrich}

\affil[a]{UC Observatories}
\affil[b]{University of California Santa Cruz}
\affil[c]{Marseille Astrophysics Laboratory}
\affil[d]{Durahm University}
\affil[e]{Wakea Consulting}
\affil[f]{King Abdullah University of Science and Technology}

\authorinfo{Further author information: (Send correspondence to Benjamin L. Gerard)\\ Benjamin L. Gerard: E-mail: blgerard@ucsc.edu}

\begin{document} 
\maketitle

\begin{abstract}
Ground-based high contrast imaging (HCI) and extreme adaptive optics (AO) technologies have advanced to the point of enabling direct detections of gas-giant exoplanets orbiting beyond the snow lines around nearby young star systems. However, leftover wavefront errors using current HCI and AO technologies, realized as “speckles” in the coronagraphic science image, still limit HCI instrument sensitivities to detecting and characterizing lower-mass, closer-in, and/or older/colder exoplanetary systems. Improving the performance of AO wavefront sensors (WFSs) and control techniques is critical to improving such HCI instrument sensitivity. Here we present three different ongoing wavefront sensing and control project developments on the Santa cruz Extreme AO Laboratory (SEAL) testbed: (1) ``multi-WFS single congugate AO (SCAO)'' using the Fast Atmospheric Self-coherent camera (SCC) Technique (FAST) and a Shack Hartmann WFS, (2) pupil chopping for focal plane wavefront sensing, first with an external amplitude modulator and then with the DM as a phase-only modulator, and (3) a laboratory demonstration of enhanced linearity with the non-modulated bright Pyramid WFS (PWFS) compared to the regular PWFS. All three topics share a common theme of multi-WFS SCAO and/or second stage AO, presenting opportunities and applications to further investigate these techniques in the future.
\end{abstract}


\section{INTRODUCTION}
\label{sec:intro}  
Adaptive Optics (AO) technology development is entering a new era, particularly in application to ground-based exoplanet and high contrast imaging (HCI) science. Current state-of-the-art extreme AO (ExAO)-fed HCI instruments enable near-infrared imaging of self-luminous exoplanets $\gtrsim$a few Jupiter masses that are on $\gtrsim$Saturn-scale orbital separations, and around nearby ($\lesssim$100 pc) young ($\lesssim$100 Myr) host stars\cite{nielsen19}. Future ExAO and HCI technology advancements are needed to enable self-luminous gas giant exoplanet imaging down to Jupiter analogs and around further away and older stars with current systems, and for future Giant Segmented Mirror Telescopes to enable habitable exoplanet imaging\cite{jensen-clem22}.

In this paper we present three new AO and HCI technologies in ongoing development in the Laboratory for AO (LAO) at UCSC, on the Santa Cruz Extreme AO Laboratory (SEAL) testbed, which is dedicated to developing and testing exoplanet imaging-focused HCI and ExAO technologies. SEAL in the context of this paper is introduced in \S\ref{sec: setup}, the three SEAL projects are then discussed in \S\ref{sec: fast}-\ref{sec: bpwfs}, and we then conclude in \S\ref{sec: conclusions}. A common theme to all three projects that we will refer to throughout this paper is the use of ``second stage AO'' and/or ``multi-wavefront sensor (WFS) single conjugate AO (SCAO).'' To clarify here, ``second stage AO'' represents a ``cascaded'' AO system configuration, where a second deformable mirror (DM) is placed downstream of the beamsplitter/dichroic to the ``first stage'' WFS path (i.e., the second stage DM is non-common path to the first stage WFS), with a separate second WFS driving that second DM. Multi-WFS SCAO instead utilizes multiple WFSs pointing at the same natural or laser guide star to control one or more common path DM(s), which we will introduce and discuss further in \S\ref{sec: multi_wfs_scao_concept}. We will also note that all laboratory results presented in this paper are planned for future publications in peer reviewed journals, and so only a concise summary of results at this time is presented in this paper.
\section{SEAL LAYOUT}
\label{sec: setup}
\begin{figure}[!h]
    \centering
    \includegraphics[width=1.0\textwidth]{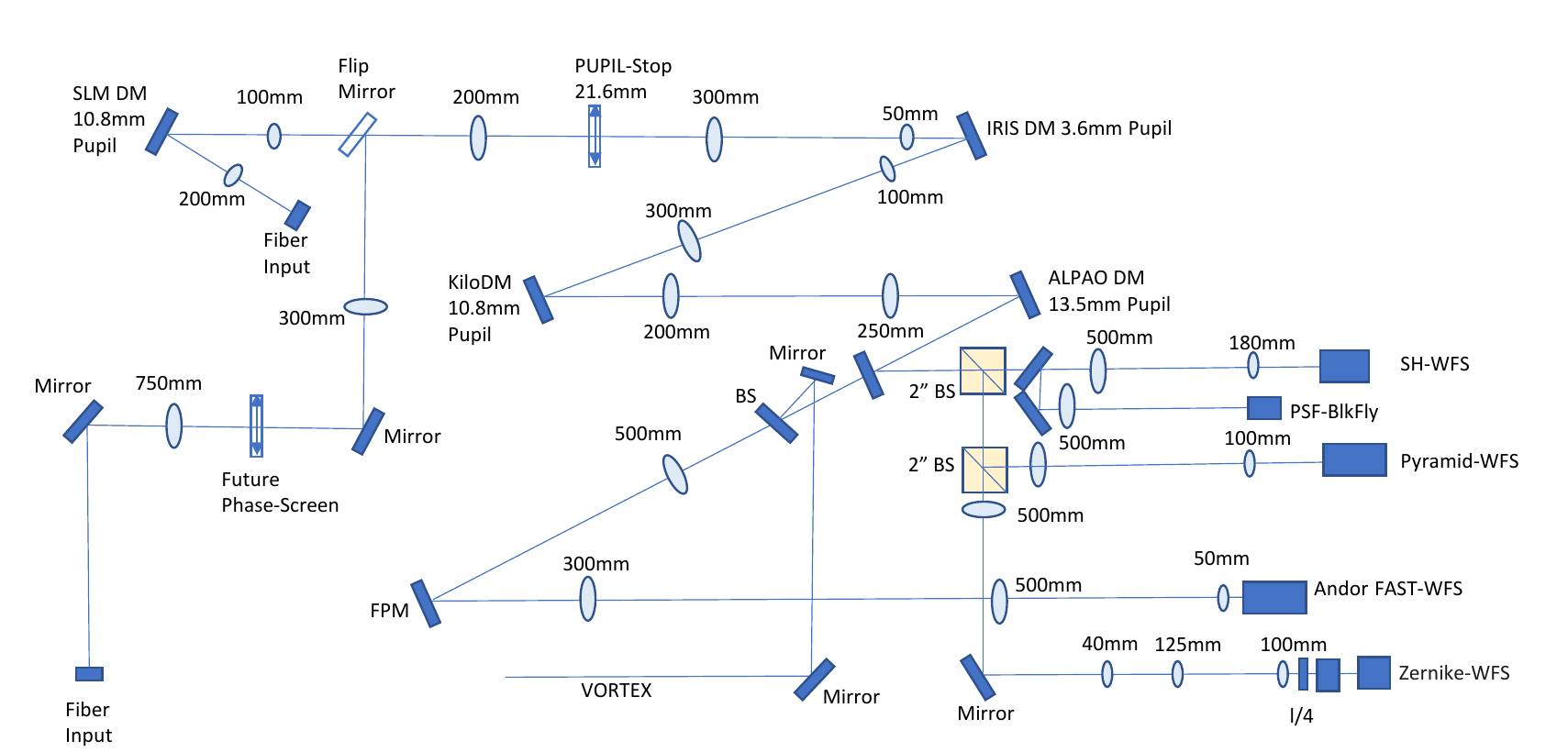}
    \caption{Layout of the SEAL testbed, previously introduced in References \citenum{jensen-clem21} and \citenum{gerard22}. This figure (image credit: D. Dillon) is also additionally presented in the following papers in this conference proceedings: References \citenum{sanchez22}, \citenum{salama22}, and \citenum{vanKooten22b}. SEAL currently operates at $\lambda=635$nm with HeNe lasers. There are currently 4 DMs (SLM, segmented Iris AO DM, low order ALPAO DM, and high order BMC kiloDM) and 6 WFS arms (SHWFS, PSF imaging camera, 3 sided Pyramid WFS, Andor FAST WFS/4 sided PyWFS, Zernike WFS, vortex coronagraph arm). Most hardware is controlled using a clone of the Keck II Pyramid WFS RTC\cite{cetre18}, with high level software written in Python.}
    \label{fig: seal_layout}
\end{figure}

Figure \ref{fig: seal_layout} illustrates the optical layout of our SEAL testbed, previously introduced in References \citenum{jensen-clem21} and \citenum{gerard22}.
Several other papers in this proceedings use SEAL: Ref. \citenum{sengupta22} for testing of Linear Quadratic Gaussian control algorithms, Ref. \citenum{sanchez22} for testing a reflective 3-sided Pyramid WFS, Ref. \citenum{salama22} for testing a vector Zernike WFS to control an IrisAO DM representing segmented telescope piston errors, and Ref. \citenum{vanKooten22b} for development of a high order spatial light modulator (SLM) for simulating laboratory atmospheric turbulence. In this paper, we will present three additional projects using SEAL, introduced below along with their relevant SEAL hardware components:
\begin{itemize}
    \item \S\ref{sec: fast}: Multi-WFS SCAO applied to FAST-SHWFS operations. This demonstration uses the Fast Atmospheric Self-coherent camera (SCC) Technique (FAST)\cite{gerard21a,gerard22} and Shack Hartmann WFS (SHWFS) sensors to control our LO (LO) ALPAO 97 actuator DM (hereafter referred to as the LODM, for low-order DM) and Boston Micromachines Corporation (BMC) high order DM (34 actuators across the beam diameter, hereafter referred to as the HODM, for high order DM, and in combination WT, for ``woofer-tweeter'').
    \item \S\ref{sec: pupil_chopping}: Focal plane wavefront sensing via a pupil chopping scheme, including first an external optical chopper device (\S\ref{sec: v1}, hereafter referred to as version 1, or v1) and second an internal DM-based approach (\S\ref{sec: v2}, hereafter v2), the latter requiring no additional hardware beyond what is already present in a conventional AO-fed imaging system. This demonstration uses the LODM, controlled for v1 by a custom optical chopper and Andor camera and for v2 by the FLIR Blackfly PSF imaging camera. 
    \item \S\ref{sec: bpwfs}: Testing the bright pyramid WFS (bPWFS), a concept introduced in Ref. \citenum{gerard21b} and now tested further with SEAL in this paper. Here we use the LODM, custom FPM, and Andor camera for bPWFS tests.
\end{itemize}
All of these above techniques present an interesting potential to further develop with the above-described multi-WFS SCAO concept, which we discuss further next in \S\ref{sec: fast}. We also note here that in this paper we utilize SEAL calibration procedures described in Ref. \citenum{gerard22}, both specifically for FAST and more generally for SEAL use (e.g., LODM And HODM voltage-to-wavefront conversion).
\section{FAST MULTI-WFS SCAO}
\label{sec: fast}
\subsection{Concept}
\label{sec: multi_wfs_scao_concept}
\begin{figure}[!h]
    \centering
    \includegraphics[width=1.0\textwidth]{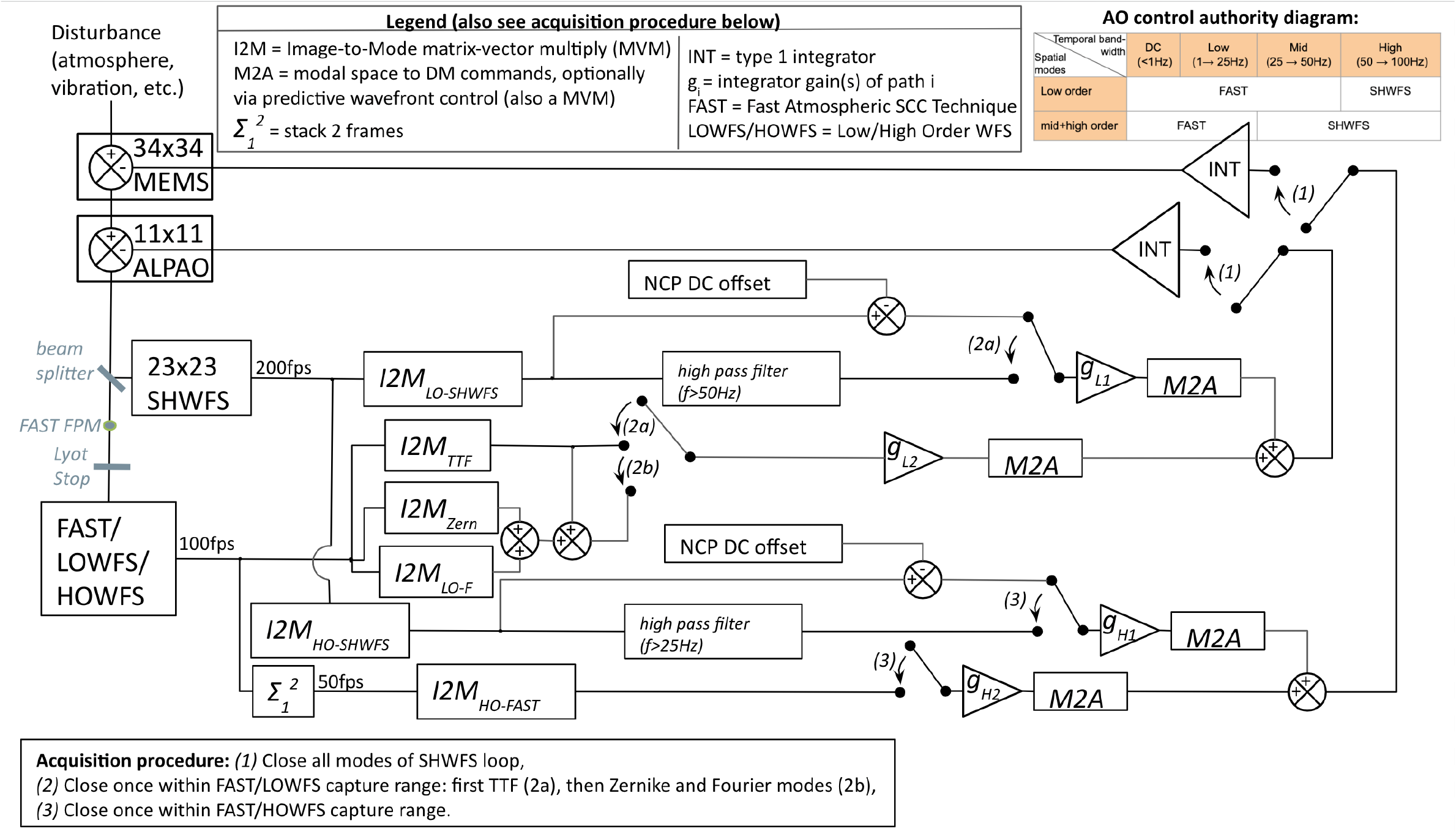}
    \caption{High-level illustration of how our SEAL SHWFS and FAST sensors, through a series of different modal groups and temporal filters, will control the LODM (11x11 ALPAO) and HODM (34x34 MEMS). See the text for more details.}
    \label{fig: fast_shwfs_diagram}
\end{figure}

Figure \ref{fig: fast_shwfs_diagram} illustrates the concept of multi-WFS SCAO, applied specifically to our SEAL FAST-SHWFS WT system. 
This diagram builds on the concept initially presented in Ref. \citenum{gerard21a}, but is now adapted specifically to the SEAL testbed. As in Ref. \citenum{gerard21a}, the main goal is to enable steady-state closed-loop control with two on-axis WFSs controlling two common-path DMs, which is done by modal groupings and temporal filters such that the main diagram with all loops closed is consistent with the upper right table in Fig. \ref{fig: fast_shwfs_diagram}, giving AO control authority to only one spatio-temporal regime for one WFS. An acquisition procedure is also illustrated in this diagram, accounting for the context that the SHWFS must first close the loop on atmospheric turbulence (step 1) and reach residual AO phases within the FAST capture range for the FAST low and high order loops to then succesively close (steps 2 and 3, respectively). Below we walk through the different real-time steps shown in Fig. \ref{fig: fast_shwfs_diagram} between given WFS image readout and sending a corresponding DM command.
\begin{enumerate}
    \item For each WFS we split modal groups between the LODM and HODM (and for the three FAST LODM sub-groups, as we found in Ref. \citenum{gerard22}).
    \item Real-time WFS frames (SHWFS center of gravity, or CoG, values, FAST $I_-$ values) are converted into modal space via a Image-to-mode (I2M) matrix-vector multiplication (MVM). 
    \item A given SHWFS modal group can then be temporally filtered (there are no FAST temporal filters here, instead giving FAST full control authority to all of it's accessible spatio-temporal regimes) on a real-time continuous buffer of modal coefficients. We have identified equation 22 in Ref. \citenum{gavel14} to enable such high pass filtering. 
    \item Modes for path $i$ are multiplied by a gain $g_i$. For modal gain optimization (MGO), this can be a vector of different optimal gain values for each mode. 
    \item Modes are converted into DM commands via another MVM. Optionally adding data-driven predictive wavefront control (PWFC), which we will discuss further below in \S\ref{sec: fast_wt_loop}, adds another MVM. If no MGO and/or PWFC is used for FAST (i.e., which does not need temporal filtering), the whole line can be reduced to a single real-time MVM.
    \item Lastly a simple type I integrator controller as shown enables closed-loop operations on both DMs. More complex controllers could be implemented here instead.
\end{enumerate}
Also note that the speeds and filter cutoffs as shown in Fig. \ref{fig: fast_shwfs_diagram} are for only a starting configuration; once we can demonstrate stable performance we can look at changing WFS frame rates and corresponding temporal filter cutoffs to simulate different guide star magnitudes, atmospheric conditions, etc.
\subsection{SHWFS-WT Loop}
\label{sec: shwfs_wt_loop}
\begin{figure}[!h]
    \centering
    \includegraphics[width=1.0\textwidth]{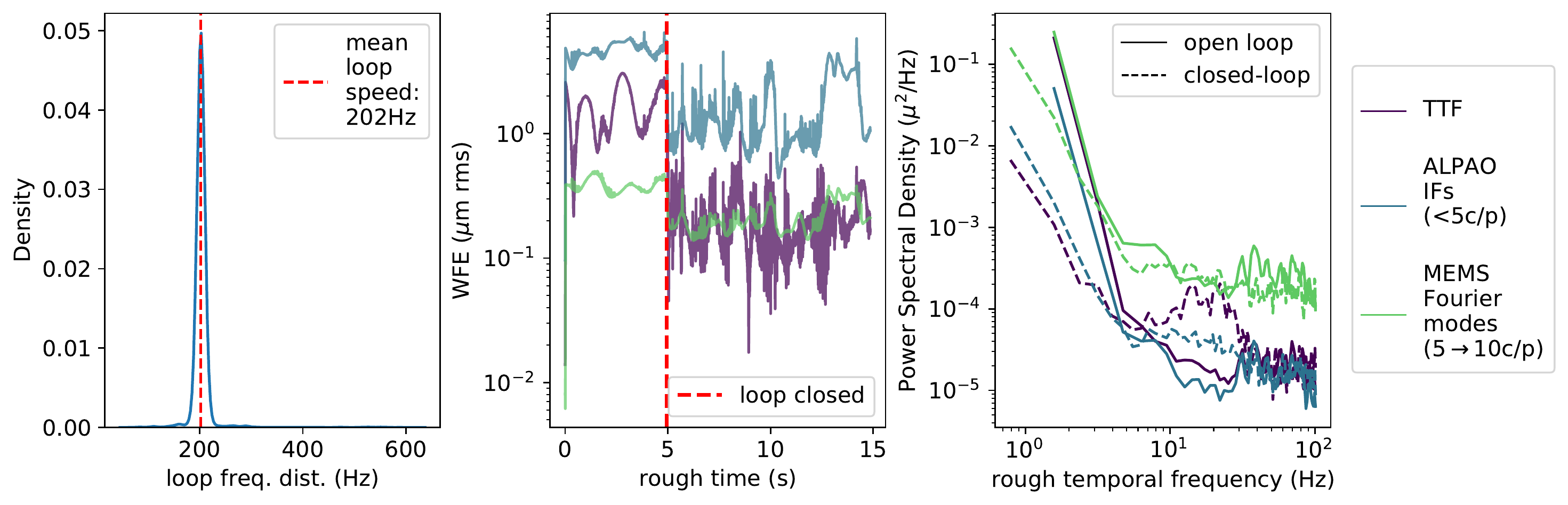}
    \caption{200 Hz SHWFS WT control of input full-scale atmospheric frozen-flow turbulence. The LODM and HODM are both used to simulate input turbulence.}
    \label{fig: shwfs_wt}
\end{figure}

Figure \ref{fig: shwfs_wt} shows telemetry results from real-time control between the SHWFS and WT DMs.
In this setup LODM modal bases include tip/tilt/focus (TTF) modes and influence functions (IFs) with TTF removed (labeled ``ALPAO IFs'' in Fig. \ref{fig: shwfs_wt}). We found this separate TTF modal group was helpful in closed-loop to enable gain and leak adjustments for a leaky integrator controller specific to only these modes. HODM modal bases are Fourier modes between 5 and 10 cycles/pupil (c/p), where the lower limit is set not to overlap with the ALPAO Nyquist limit and the upper limit is set below the 23$\times$23 SHWFS lenslet array sampling. Although this setup under-samples HODM-SHWFS Fried geometry, this 23$\times$23 subaperture modes is the only configuration available for this Thorlabs WFS-20 in high speed mode, which enables on-chip CoG processing to provide a stable and high frame-rate readout at $\sim$200 Hz (with Thorlabs benchmarking possible speeds up to 1 kHz), as showing in the left panel of Fig. \ref{fig: shwfs_wt}.

We generate input atmospheric turbulence using the LODM and HODM via the WT framework presented in Ref. \citenum{gavel14}, in particular that the spatial frequency cutoff to switch from LODM to HODM turbulence is set well below the LODM Nyquist limit. Note that this cutoff for injected DM-based turbulence is different from the LODM control frequency cutoff, the latter instead controlling out to the LODM Nyquist limit. We use an offline pre-calibrated interaction matrix (IM) between LODM LO injected turbulence modes, HODM injected turbulence modes, and SHWFS slopes in order to project a full power law input atmospheric turbulence in HODM space onto LODM and HODM commands to be applied in real-time via a single MVM.

As shown in Fig. \ref{fig: shwfs_wt}, we are still adjusting and tuning this SHWFS-WT loop, as AO residual phases are not yet at the Extreme AO (ExAO) levels we would like. A 5 m/s frozen flow single-layer wind speed is used here, but with higher wind speeds planned in the future with more faster loop speeds and a more optimized control framework (e.g., implementing MGO and/or PWFC). However, our first attempts at SHWFS-WT MGO showed that open and closed loop PSDs do not match the theoretical rejection trasnfer function given the measured loop speed. Since this measured loop speed is for our Python serial implementation (i.e., measured times include WFS image acquisition, computational lag, and applying DM commands), it is hard to identify what factors might be causing limited performance gains. We have already started developing a low-level C-based parallelized loop code that should help improve stability and performance before integrating this loop with the FAST WT loop.
\subsection{FAST WT Loop}
\label{sec: fast_wt_loop}
\begin{figure}[!h]
    \centering
    \includegraphics[width=1.0\textwidth]{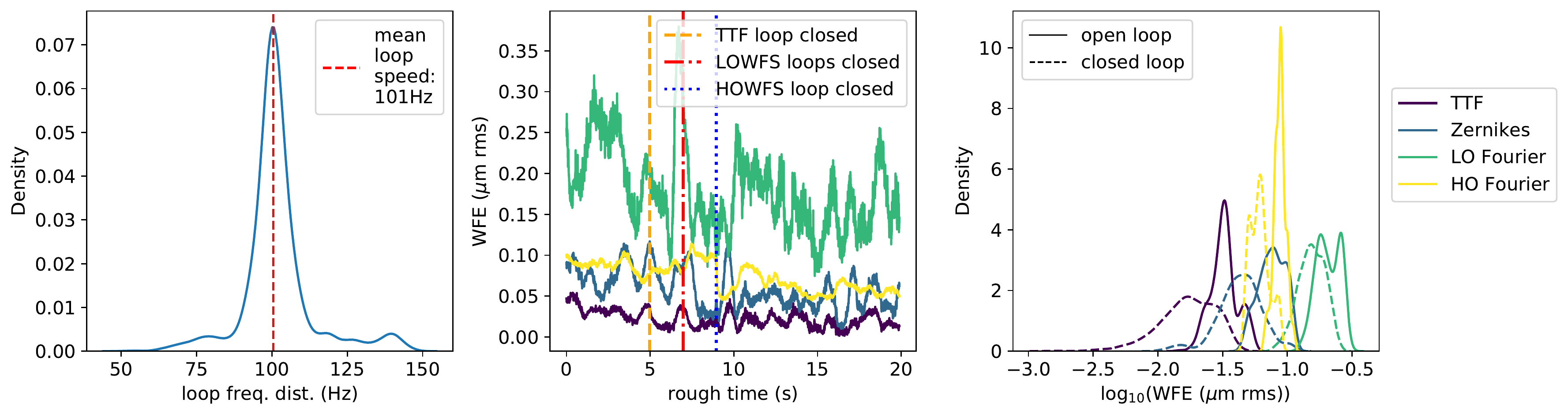}
    \caption{100 Hz FAST WT control in input frozen flow AO-residual atmospheric turbulence. The HODM is used to simulate input turbulence. A 50 Hz loop speed version of this figure was initially published in Ref. \citenum{gerard22}.}
    \label{fig: fast_wt}
\end{figure}

Figure \ref{fig: fast_wt} shows closed-loop telemetry results for FAST running real-time correction on HODM-injected AO residual turbulence.
The panels are similar to Fig. \ref{fig: shwfs_wt}, but with slightly different modal groups and acquisition procedures. A 1 m/s frozen flow wind speed is used, with planned higher wind speeds accessible by higher frame rates and/or more advanced real-time control algorithms. These results are consistent with the FAST findings in Ref. \citenum{gerard22}, where we demonstrated how four different modal groups enabled good linearity and closed-loop performance: TTF, LO Zernike modes (labeled ``Zernikes''), LO Fourier modes (labeled ``LO Fourier,'' still diffracting sine spots within the FAST focal plane mask inner working angle), and high order Fourier modes (labeled ``HO Fourier,'' using the conventional modal basis to generate a half dark hole, or DH). The former three modes are controlled with the LODM, where as the latter high order Fourier modal group is controlled with the HODM. We also implement the same LO acquisition procedure as in Ref. \citenum{gerard22} and as in Fig. \ref{fig: fast_shwfs_diagram}: first closing TTF, then closing LODM Zernike and Fourier modes, and lastly closing HODM Fourier modes, which we found was necessary to enable stable closed-loop DHs over time. Again, similar to \S\ref{sec: shwfs_wt_loop}, we are still tuning parameters for optimal closed-loop, including ongoing developments to migrate our Python real-time control software into C in a format consistent with Fig. \ref{fig: fast_shwfs_diagram}.

\begin{figure}[!h]
    \centering
    \includegraphics[width=0.48\textwidth]{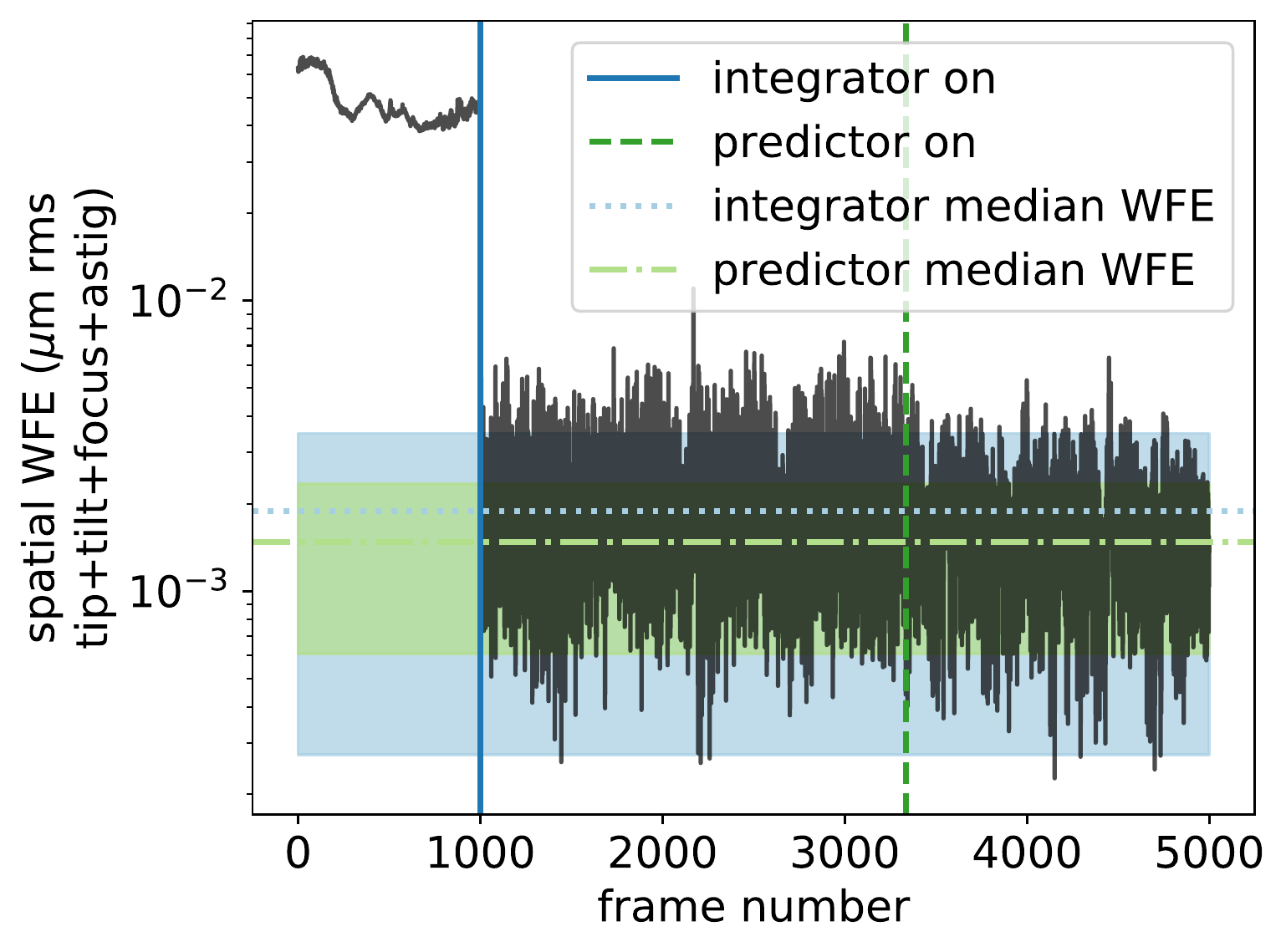}
    \includegraphics[width=0.5\textwidth]{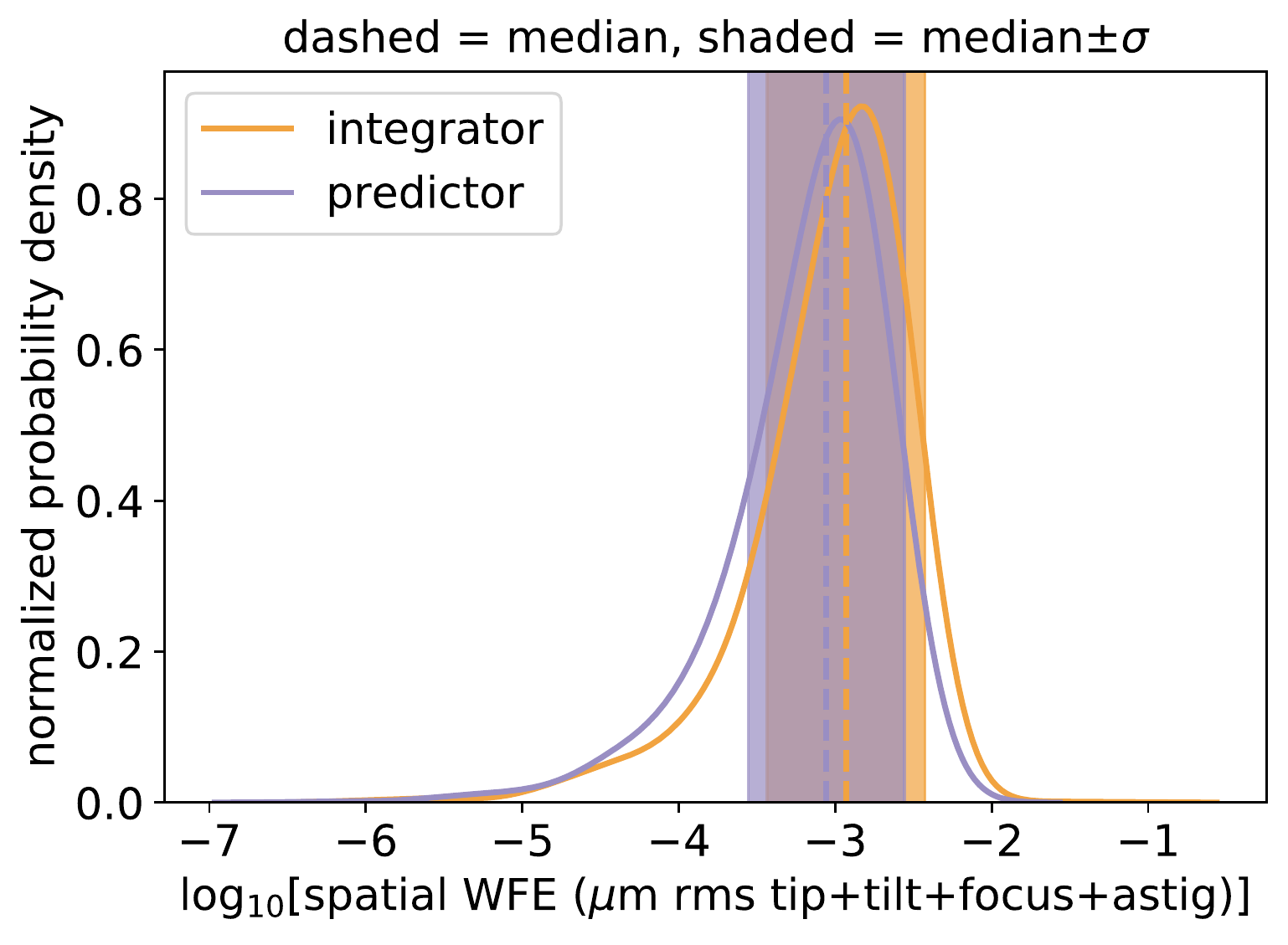}
    (a) \hspace{250 pt} (b)
    \caption{FAST PWFC for LO Zernike modes of frozen flow AO residual turbulence, showing the time domain version in (a), running at $\sim$50 Hz and the time-collapsed histograms in (b). The shaded regions in each panel represent the standard deviation around the mean represented by the line of the same color.}
    \label{fig: fast_lo_pwfc}
\end{figure}

We have also demonstrated real-time FAST PWFC, to our knowledge a first of it's kind for focal plane WFSs, shown in Fig. \ref{fig: fast_lo_pwfc}.
Here we are taking a similar approach to what our UCSC LAO group is currently testing at W. M. Keck Observatory\cite{vanKooten22a}, we implement PWFC using Empirical Orthogonal Functions (EOF)\cite{guyon17} method. Using EOF we predict the closed-loop low order modal coefficients (note that this is different from the original EOF implementation that requires a pseudo open loop reconstruction) by storing up modal coefficients to generate a predictive filter. The filter is then used in a MVM to subsequently predict the FAST coefficients over the system delay, represented by the ``M2A'' boxes in Fig. \ref{fig: fast_shwfs_diagram}. In Fig. \ref{fig: fast_lo_pwfc}, we show preliminary results using EOF with FAST for LO modes. The distributions of residual wavefront error (WFE) over time indicate an improvement of 1.3x when EOF (i.e., PWFC) is used, with a WFE standard deviation improvement of 1.9x (the latter indicating increased stability). We plan to further improve the performance of FAST PWFC by optimizing the predictor order and the amount of training data.

\section{PUPIL CHOPPING FOR FOCAL PLANE WAVEFRONT SENSING}
\label{sec: pupil_chopping}
We present two versions of the pupil chopping concept here: the first version (hereafter v1) in \S\ref{sec: v1} uses an external optical chopping device. The second version (hereafter v2) in \S\ref{sec: v2} is a newer concept that continues to be developed in simulation and tested in the lab (project lead: co-I J. Perez-Soto), and so we present preliminary results in \S\ref{sec: v2}.
\subsection{v1: External Optical Chopping}
\label{sec: v1}
\begin{figure}[!h]
\includegraphics[width=0.5\textwidth]{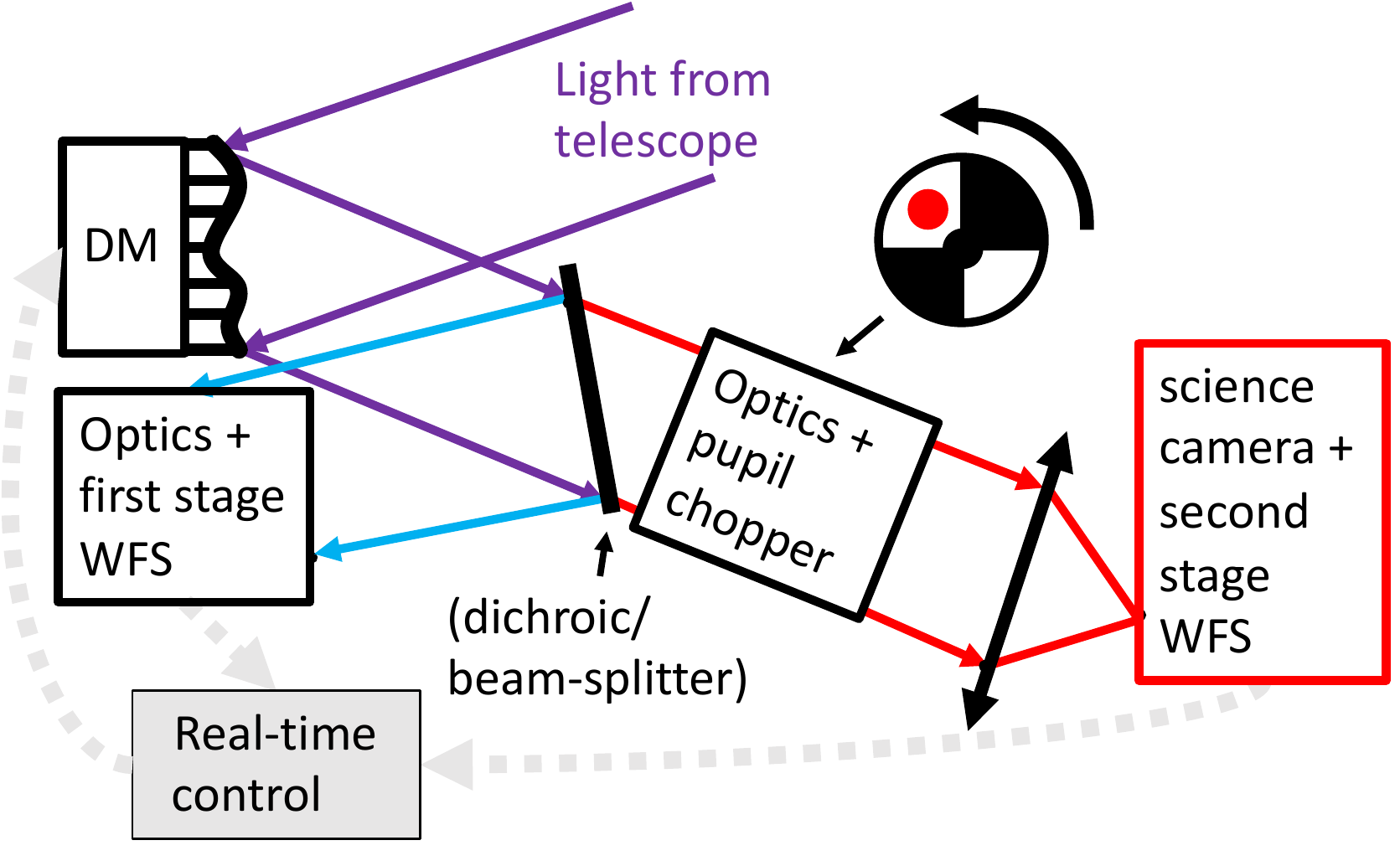}
\includegraphics[width=0.5\textwidth]{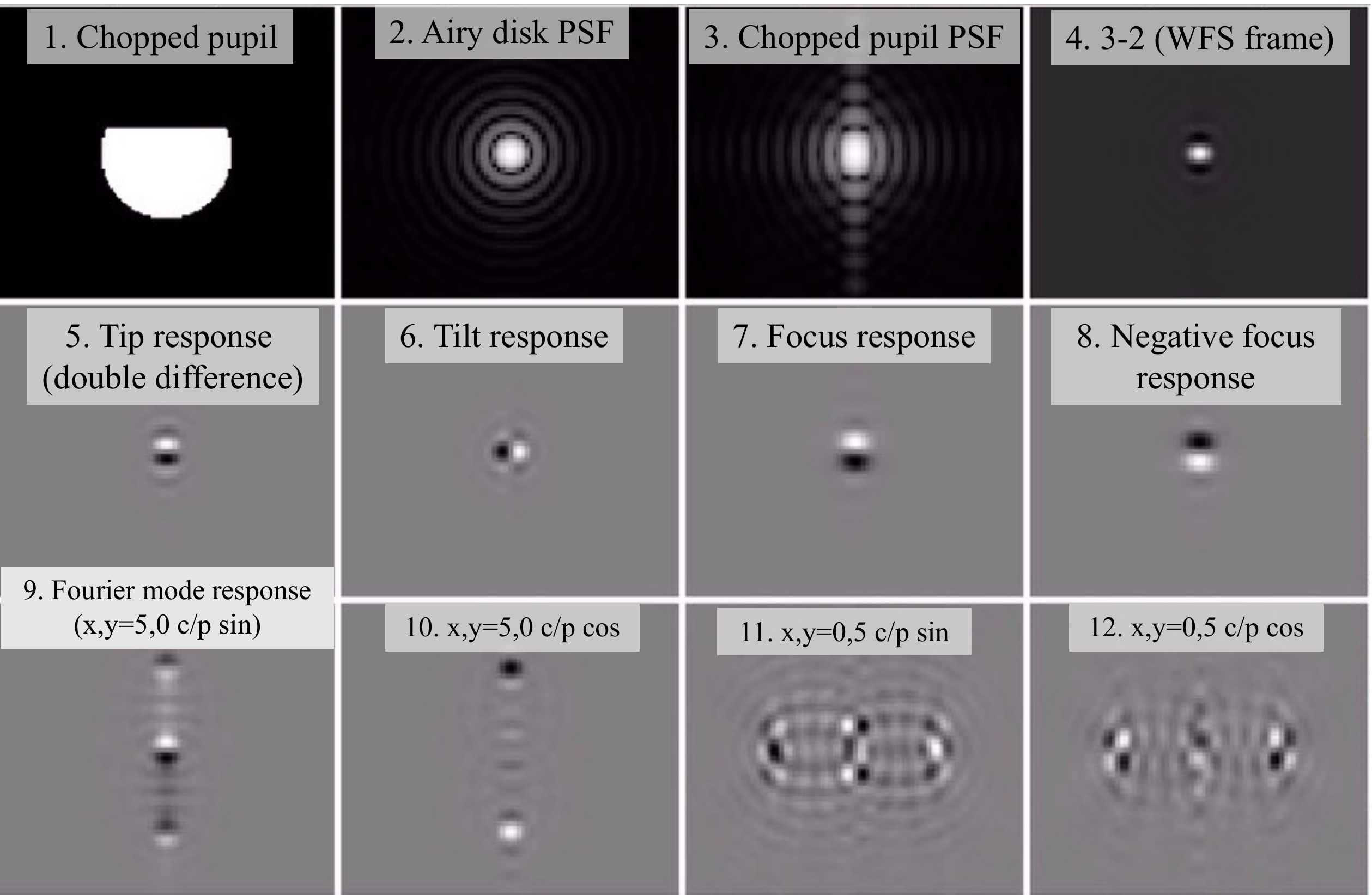}
\hspace*{150pt} (a) \hspace{200pt} (b) 
\caption{Illustration of the pupil chopping concept, both from a geometric (a) and Fourier optics (b) perspective. In (b), all focal plane images are shown with the same field of view. Panels 2-3 are shown on a log scale; all other panels are shown on a linear scale. Zernike and Fourier modes in panels 5-12 all use 1 nm amplitudes. Panels 7-12 indicate that pupil chopping for focal plane wavefront sensing can resolve the sign ambiguity of focus and the relative phase of Fourier modes, which is not possible with a single PSF image.}
\label{fig: v1_concept}
\end{figure}

Figure \ref{fig: v1_concept} illustrates the concept of focal plane wavefront sensing with a pupil plane optical chopper.
In short, two focal plane images are required to enable a WFS measurement: one with the chopper blade partially blocking the pupil, and one with the pupil unblocked; the latter can be used for science while the former is limited to wavefront sensing. A ``reference'' WFS frame is saved before on-sky operations and then used analogously to the concept of on-sky reference slopes for other pupil plane WFSs, enabling a real-time WFS frame to in principle flatten the wavefront to the level of the pre-calibrated reference WFS frame. It should be noted that this technique is complementary to pupil amplitude diversity-based absolute phase retrieval methods, such as the asymmetric pupil Fourier WFS\cite{martinache13}. This technique enables a linear-least squares reconstruction, optimal for high speed wavefront control of AO residuals but not able to flatten the absolute phase below a pre-calibrated best flat (e.g., due to evolving on-sky quasi-static aberrations), whereas phase retrieval methods can use the chopped pupil PSF image to recover the absolute phase to track such quasi-static effects, but typically requiring non-linear iterative algorithms that cannot be run at high speed.

Our SEAL laboratory setup for v1 pupil chopping presented in this subsection uses the LODM and Andor camera as previously descrived, and additionally a Thorlabs MC2000B optical chopper with blade MC1F10, and a digital signal processing controller and custom electrical doubler so that the Andor camera readout is synchronized with the optical chopper (with the chopper as the leader and the Andor camera as the follower\footnote{Although the AO field has historically adopted ``master and slave'' terminology from the computer science field to refer to the chopper and camera in this configuration, respectively, it is now clear that such discriminatory terminology should no longer be used. Following the computer science field, instead we will refer to such configuration as ``leader and follower'' throughout this paper, encouraging other members of the AO field to do the same.}) to produce a chopped and un-chopped pupil every other frame, respectively.

\begin{figure}[!h]
    \centering
    \includegraphics[width=0.48\textwidth]{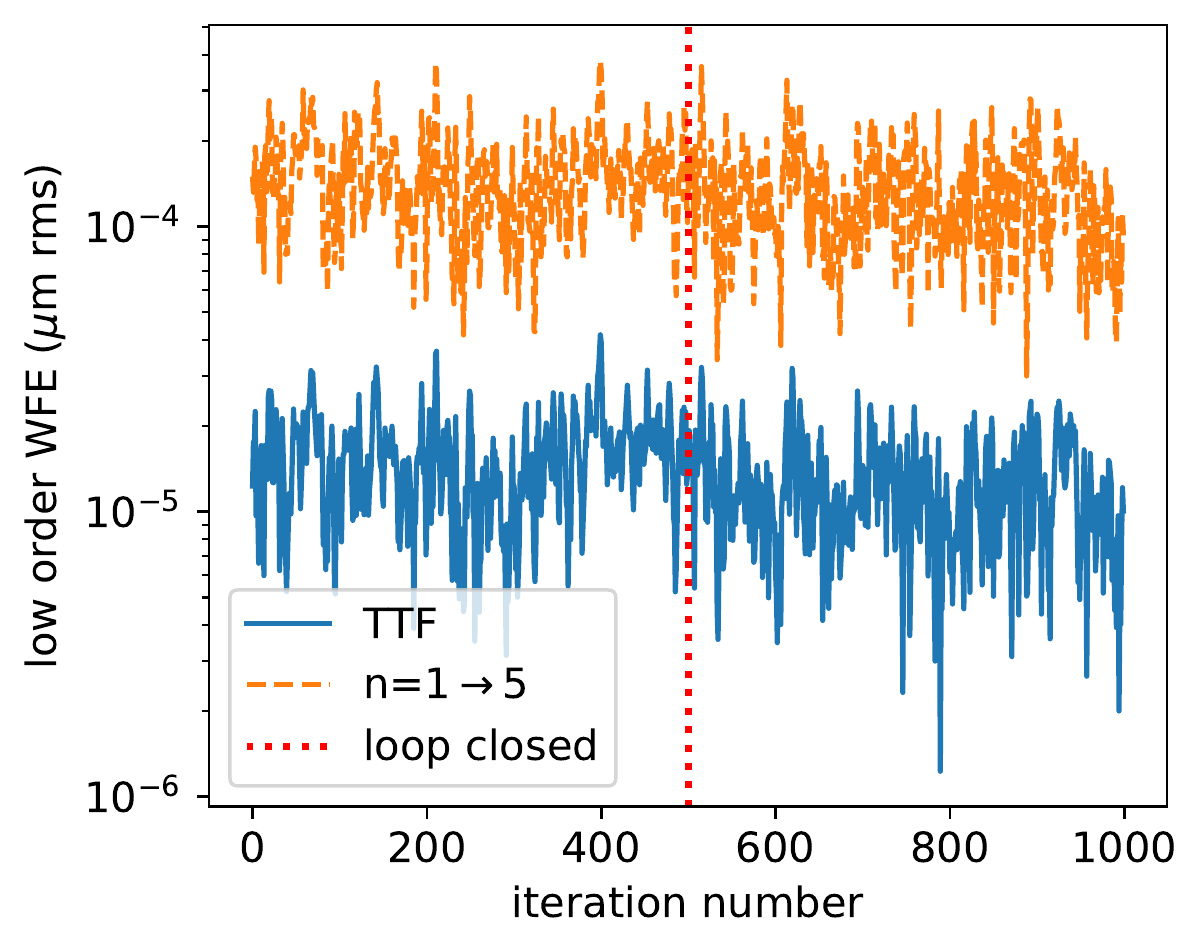}
        \includegraphics[width=0.48\textwidth]{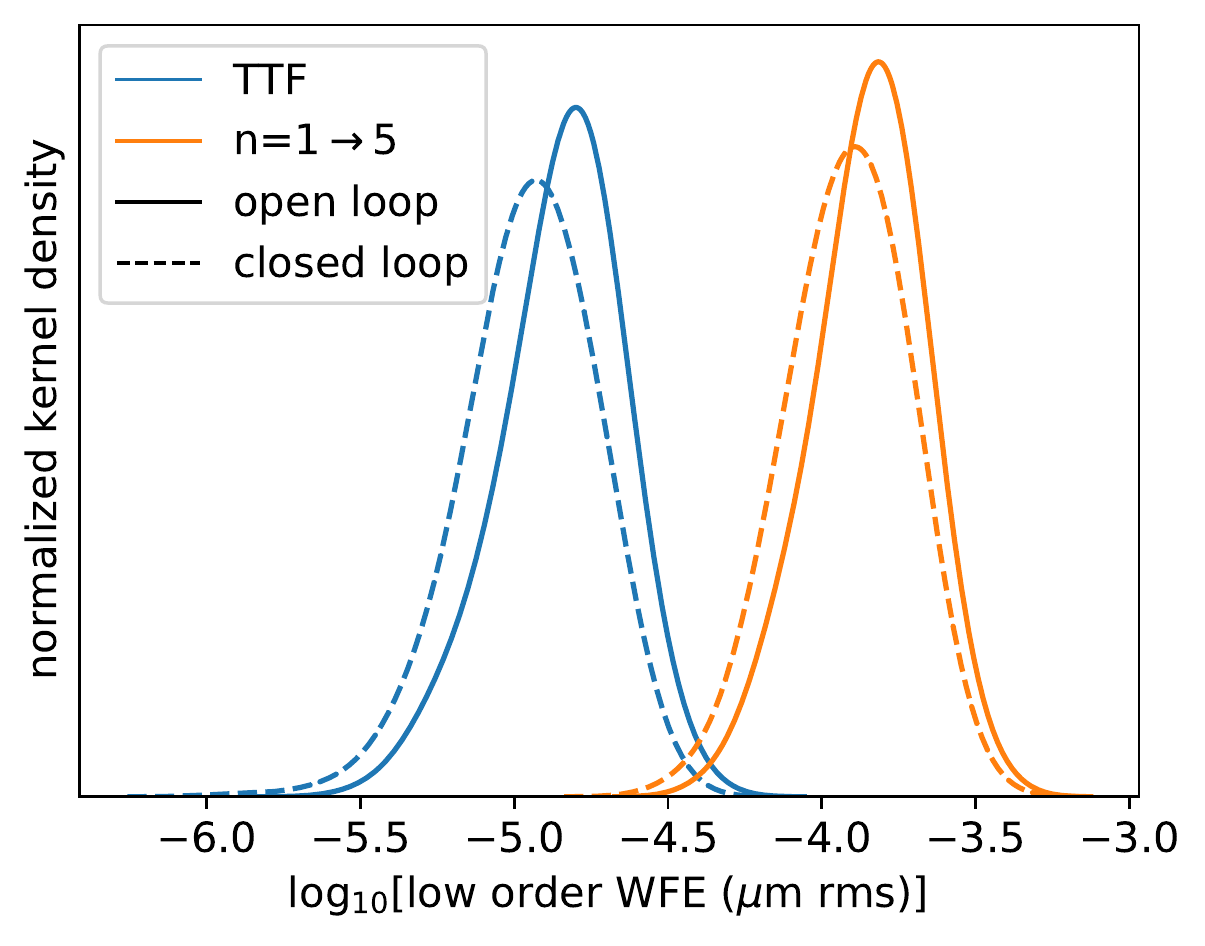}
    \\
    \hspace{-10pt} (a)  \hspace{200pt} (b)
    \caption{Real-time correction of on-air wavefront error (at a frame rate of about 25 Hz), showing both tip/tilt/focus (TTF) and all controlled modes ($n=1\rightarrow5$), both in the time domain (a) and by WFE distribution comparing the open and closed loop sequences (b; i.e., respectively to the left and right of the dotted line in panel a). Note that the y axis in panel (a) and x axis in panel (b) are not correctly normalized; only the relative reduction between open and closed loop WFE should be considered for the purposes of this paper.}
    \label{fig: v1_on_air}
\end{figure}

Using a Zernike modal basis of up to five radial orders (limited to higher orders by phase jitter of the chopper blade in our lab setup), on-air closed-loop results are shown in Fig. \ref{fig: v1_on_air}.
Although the highly stabilized environment from our granite optical table allows only a moderate gain, Fig. \ref{fig: v1_on_air} clearly demonstrates a closed-loop WFE reduction relative to open loop for all controlled modes with only on-air temporal disturbances present, illustrating the potential of this technique to stabilize quasi-static temporal disturbances on-sky (e.g., thermal variations, flexure, and/or other sources of slow beam wander). Note these results help address concerns  that the spinning chopper wheel would be generating additional on-air turbulence, since telemetry from closing the loop on-air clearly shows improvement compared to open-loop.

\begin{figure}[!h]
    \centering
    \includegraphics[width=1.0\textwidth]{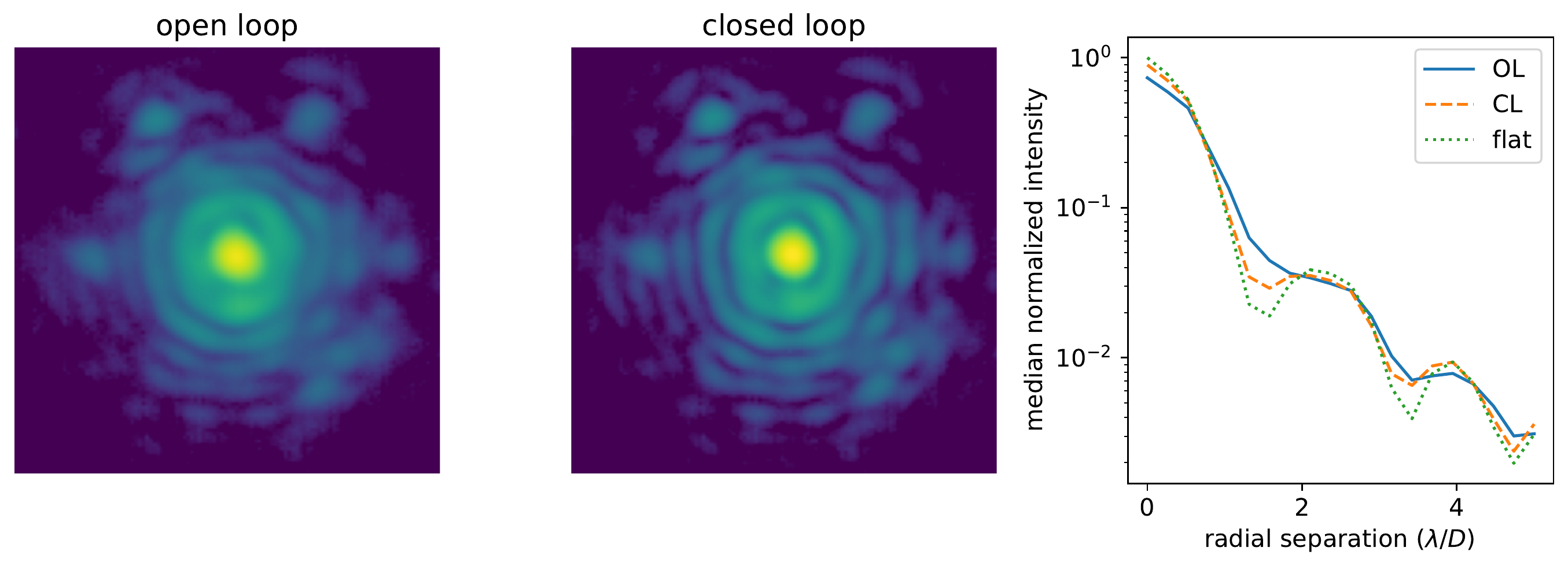}
    \caption{Open and closed loop image stacks of un-chopped images (left and middle panels) and corresponding radial intensity profile (right panel) for frozen flow 10 m/s AO residual turbulence applied on the DM in real time at $\sim$50 Hz. The right panel shows a $\sim$20\% Strehl improvement by closing the loop on AO residuals.}
    \label{fig: v1_on_turb}
\end{figure}
Building on these on-air results, Fig. \ref{fig: v1_on_turb} shows un-chopped PSF images and a corresponding radial profile for LODM-injected AO residual turbulence with and without closing the loop on that turbulence, clearly illustrating the potential of this technique for second stage AO and/or multi-WFS SCAO. In addition to the above-mentioned chopper blade phase jitter limiting our ability to measure and control Zernike modes beyond 5 radial orders (for which a custom chopper solution could reduce sufficiently to enable higher order control, such as Ref. \citenum{johnson22}), a more fundamental complexity is that an additional optical chopper hardware component must be designed and installed into instruments for this technique to work, increasing project cost and complexity. Science duty cycle with this v1 technique is also limited to, at most 50\%. Instead, v2, discussed next in \S\ref{sec: v2}, improves on all of these issues (i.e., removes the phase jitter problem, does not require additional optical chopper hardware, and increases the science duty cycle).
\subsection{v2: DM-based Chopping}
\label{sec: v2}
\begin{figure}[!h]
    \centering
    \includegraphics[width=1.0\textwidth]{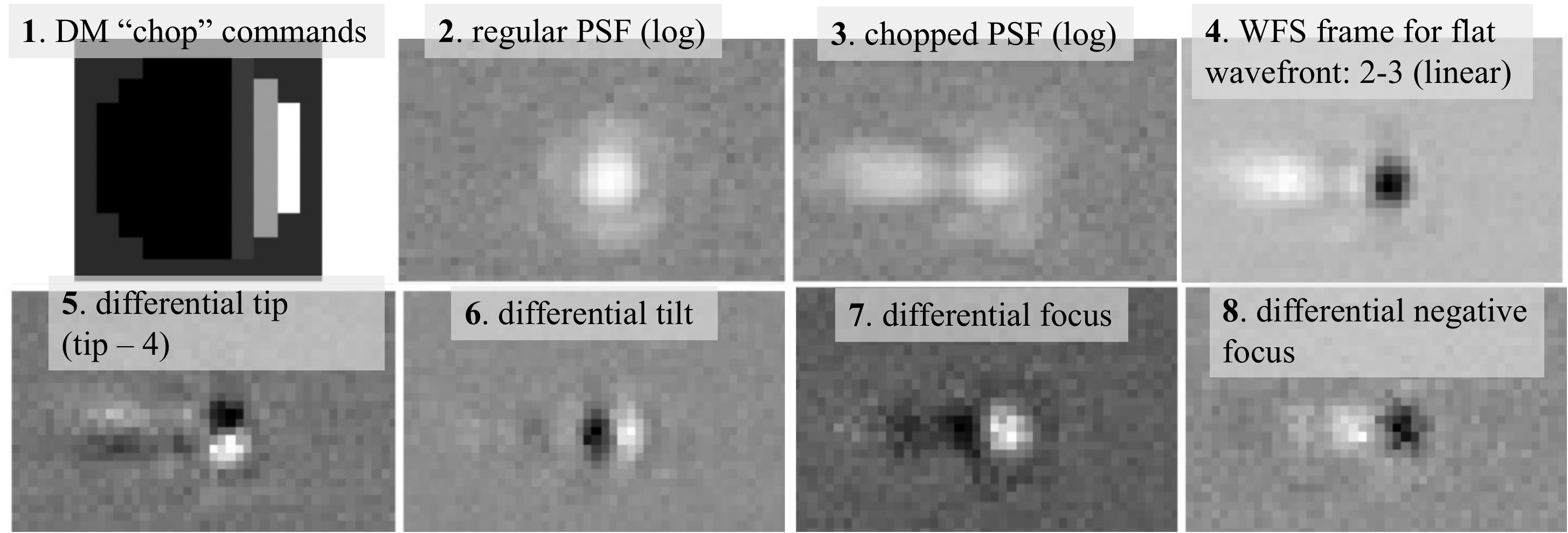}
    \caption{DM-based pupil chopping concept illustration, using SEAL testbed images, clearly demonstrating the potential of this to resolve the sign of focus and other common degenerates present from a single PSF image, but still enable high duty cycle science imaging. }
    \label{fig: v2_concept_seal}
\end{figure}

Instead of using a physical optical chopper device, v2 of this approach uses the DM of an existing AO system to apply a local tilt to a fraction of the illuminated pupil. PSF images are recorded where this DM ``chop'' commands are added to all other DM commands every other frame. Fig. \ref{fig: v2_concept_seal} illustrates this concept using laboratory images from SEAL.
At the moment this local tilt uses a peak-to-valley stroke of $\sim$1$\mu$m with our LODM, but we are still investigating whether lower stroke options (e.g., enabling HODM-based pupil chopping without saturating available DM stroke) are compatible with this technique. Although Fig. \ref{fig: v2_concept_seal} only shows low order wavefront modes, this technique is broadly applicable to wavefront sensing and control of spatial orders out to the DM Nyquist limit, and by being a focal plane wavefront sensor it benefits from natural wavefront spatial filtering properties, enabling binary masks on the focal plane image to act as a natural anti-aliasing filter and minimize non-linear cross talk between modes (e.g., as demonstrated for FAST in Ref. \citenum{gerard22}). Using the focal plane image and such binary masks for wavefront sensing with this technique also distinguishes this technique from the differential optical transfer function method\cite{codona13}.

\begin{figure}[!h]
    \centering
    \includegraphics[width=1.0\textwidth]{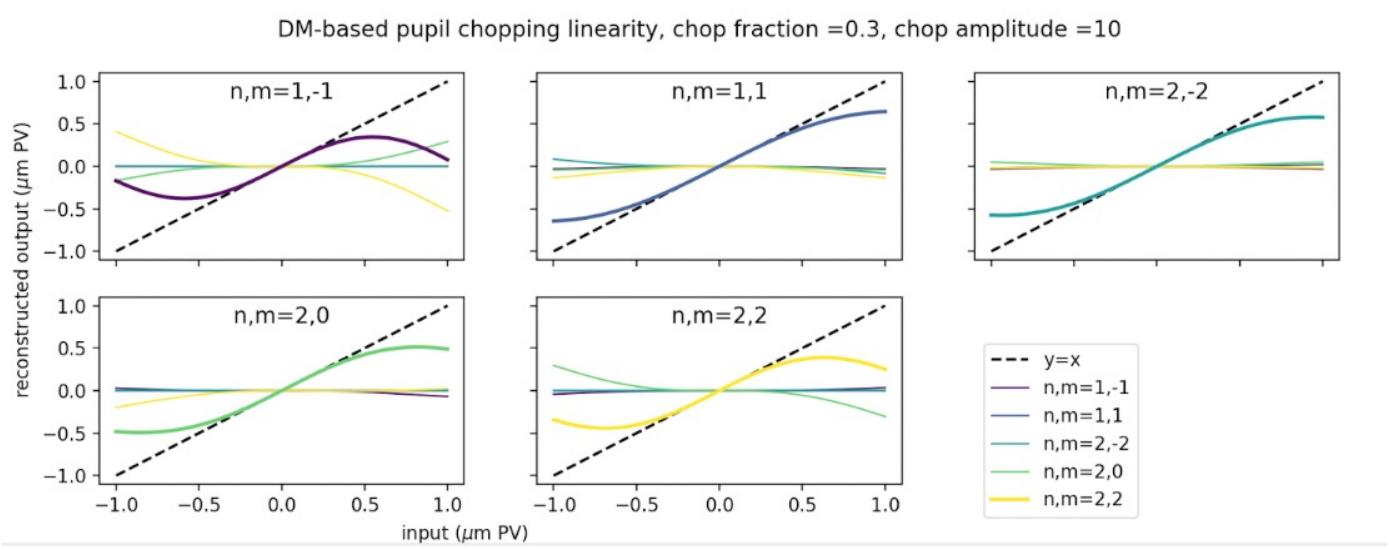}
    \caption{LO linearity simulations for our proposed DM-based pupil chopping technique, assuming $\lambda$=1.6$\mu$m to show peak-to-valley (PV) units on both axes.}
    \label{fig: v2_linearity}
\end{figure}

Figure \ref{fig: v2_linearity} shows preliminary DM-based pupil chopping linearity simulations for low order modes.
As expected from Fig. \ref{fig: v2_concept_seal}, LO modes show excellent linearity for phases $\lesssim$1 rad rms, showing applicability to correct AO residual phases but not open loop atmospheric turbulence. Further simulations and SEAL experiments with this DM-based pupil chopping technique are ongoing.

The hardware setup of this technique is extremely simple, requiring only a PSF imaging camera in addition to existing AO system hardware, making this compatible with most current and future AO systems. A high-speed PSF imaging camera would enable second stage and/or multi-WFS SCAO correction of AO residuals, but even a slower readout camera could enable on-sky stabilization of quasi-static wavefront errors. This technique also has broad applicability to other AO fields (e.g., laser guide stars, off-axis stars, crowded star fields, extended sources); as long as a camera images the focal plane of the intended light source for guiding and the DM is conjugated to the pupil plane, the principles of this technique remain unchanged. Increasing the science duty cycle above 50\% is also feasible with this setup (e.g., if the average PSF camera readout is at 1 kHz, if the guide star is bright enough the chopped pupil frame can be a 0.1 ms exposure and the un-chopped pupil science frame can be 0.9 ms, enabling a 90\% science duty cycle). However, a multi-WFS SCAO approach with this technique means that the other AO WFS sees the local DM tilt every other frame. Possible solutions to this are to (1) remove this local tilt mode from the control basis for the first stage WFS, (2) drop every frame when the DM has this local tilt applied (which if operating, e.g., at a 90\% duty cycle should yield a minimal performance loss), or (3) use this technique only for second stage AO, meaning that only a second DM that is not common path to the first stage WFS is driven by this DM-based pupil chopping technique. All options should be investigated further to more quantitatively understand the benefits and disadvantages of each approach.
\section{TESTING THE BRIGHT PYRMAID WFS}
\label{sec: bpwfs}
\begin{figure}[!h]
    \centering
    \includegraphics[width=0.5\textwidth]{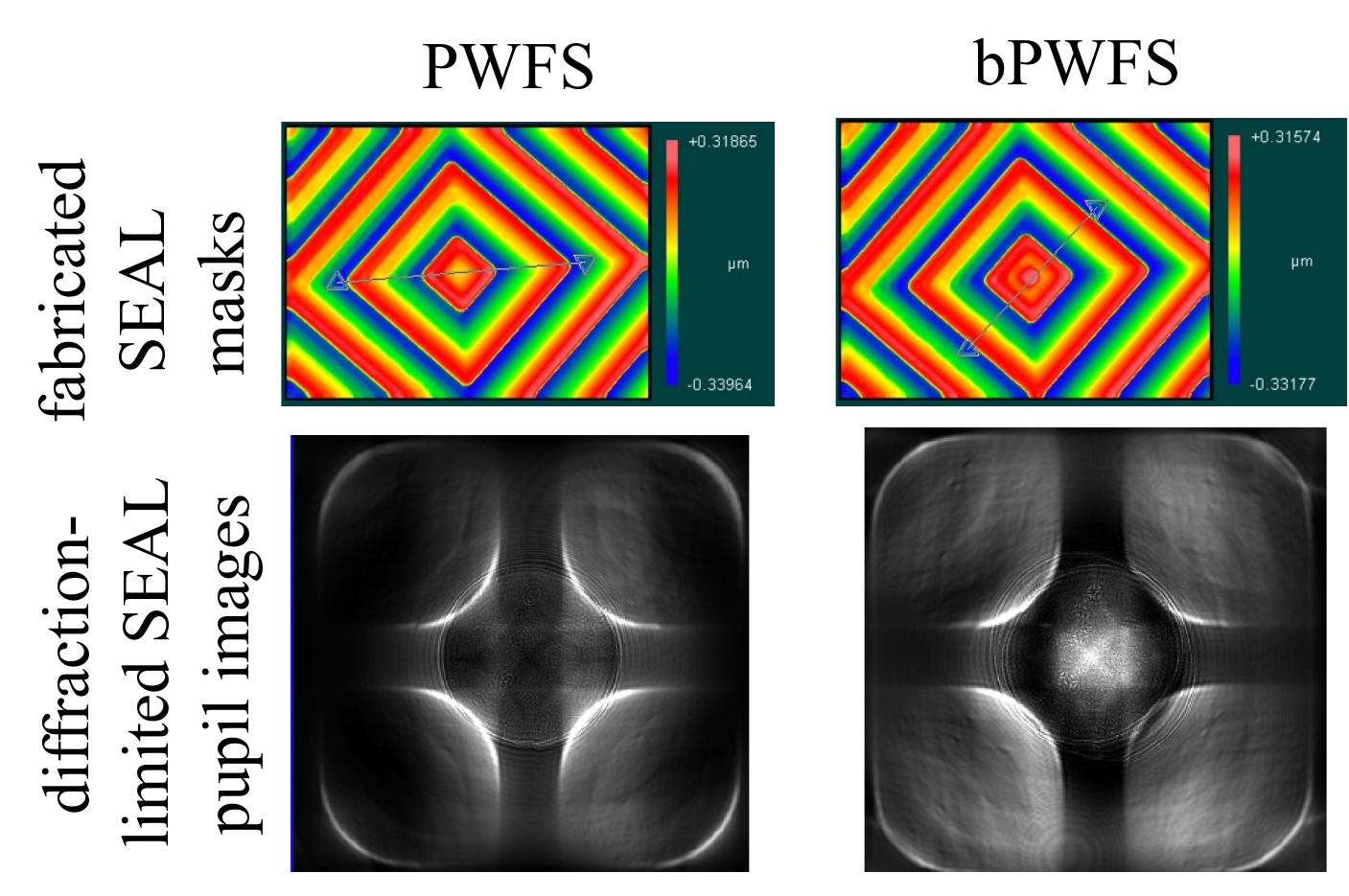}
    \includegraphics[width=0.46\textwidth]{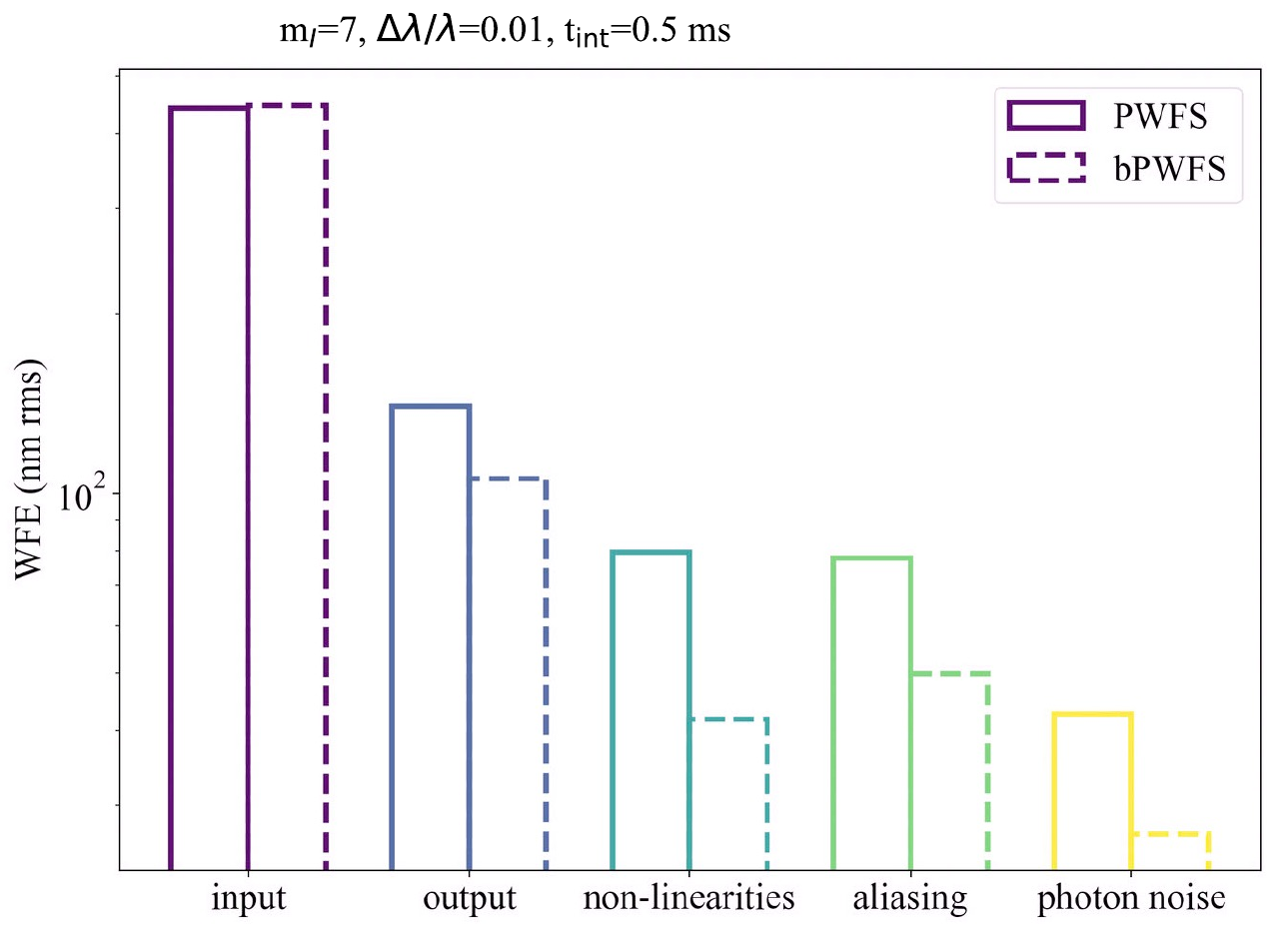}
    \\
    (a) \hspace{175pt} (b)
    \caption{Illustration of the bPWFS concept, first introduced in Ref. \citenum{gerard21b}. In panel a, we show fabricated masks (i.e., Zygo interferometer-measured depth profiles) from KAUST and corresponding pupil images from SEAL after aligning these masks, illustrating that the non-modulated bPWFS diffracts more light into the pupil footprints than the non-modulated PWFS (the PWFS and bPWFS pupil images are both normalized to the same scale). The benefits of that light redistribution are illustrated in panel b, showing simulations (adapted from Ref. \citenum{gerard21b} to integrate over spatial frequencies less than 16 c/p for an assumed 32 actuator diameter DM), showing that for input AO residual phases, the bPWFS vs. PWFS output improves the non-linearity, aliasing, and photon noise error budget components.} 
    \label{fig: bpwfs_concept}
\end{figure}

Lastly, we have recently been using SEAL to further test and develop the bright Pyramid WFS (bPWFS) concept, initially presented in Ref. \citenum{gerard21b} and illustrated here in Fig. \ref{fig: bpwfs_concept}.
In this paper, we present further testing of the bPWFS concept on SEAL, using custom-made reflective 16-level scalar Aluminum masks fabricated by the Computational Imaging group at King Abdullah University of Science and Technology (KAUST). This fabrication process, reaching $\sim$3 micron spatial resolution, $\sim$few nm-level depth precision, but limited to $\sim$800 nm dynamic range (hence the need to phase wrap the PWFS tilt angles), is described in detail in Ref. \citenum{fu21}. Despite the need for phase wrapping, which is inherently monochromatic, the high depth precision and spatial resolution could potentially outweigh chromaticity effects from phase wrapping to improve PWFS quality compared to existing fabrication techniques, and would be interesting to consider for applications in other PWFS systems. In Fig. \ref{fig: bpwfs_concept}, our high quality KAUST PWFS and bPWFS masks made specifically for SEAL clearly show more light within bPWFS vs. PWFS pupil footprints, illustrating the potential to demonstrate enhanced performance in the lab as simulations suggest.

Our first approach was to consider a closed-loop bootstrapping procedure where a modulated bPWFS is first used to close the loop on atmospheric turbulence and then modulation is switched off to see the non-modulated bPWFS gains. However, we found in simulations, both for the PWFS and bPWFS, that non-linearities dominate the error budget when modulation is switched off and as such performance immediately degrades. Instead, in this paper we only investigate non-modulated linearities for the two WFSs, keeping in mind that future solutions will be needed to integrate this non-modulated bPWFS performance gain into real AO systems, such as a second stage bPWFS, multi-WFS SCAO approach, and/or non-linear bPWFS reconstruction algorithms.

\begin{figure}[!h]
    \centering
    \includegraphics[width=1.0\textwidth]{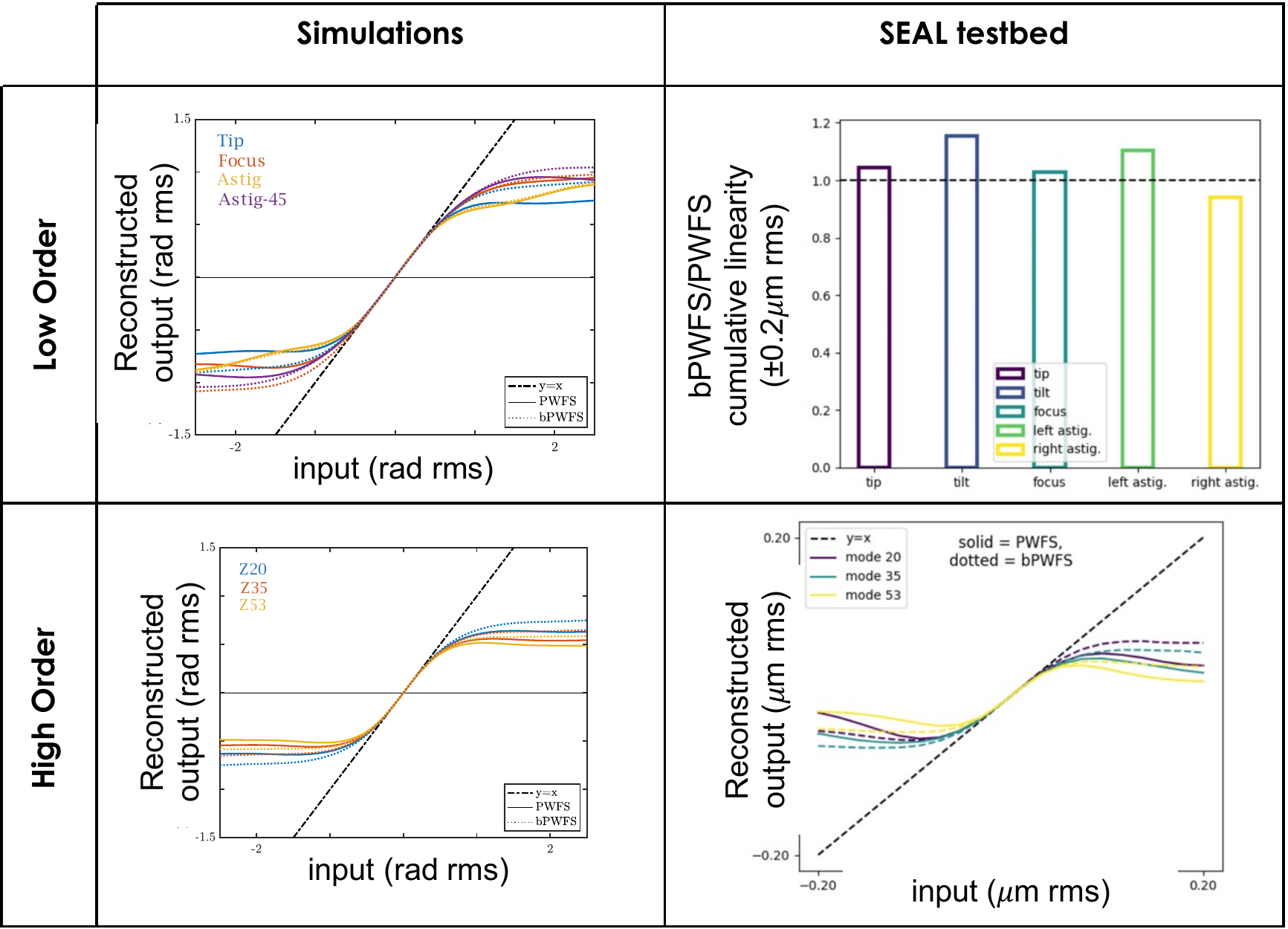}
    \caption{OOMAO simulations and corroborating SEAL data (left and right columns, respectively) showing the non-modulated bPWFS is more linear (i.e., both linear and capture range) than the regular PWFS for both low and high order Zernike modes (top and bottom rows, respectively). Due to alignment issues, SEAL LO linearity compares the integral of the absolute value of linearity curves for different LO Zernike modes, showing the ratio of this integral for the bPWFS relative to the PWFS (i.e., a number greater than one indicates enhanced bPWFS linearity). This cumulative linearity metric works for our lab setup where there is only small misalignment between the PWFS and bPWFS masks, but it would not be representative for larger alignment errors where the probed linear range does not reach the saturated region on both positive and negative ends of a given linearity curve.}
    \label{fig: bpwfs_linearity}
\end{figure}

Figure \ref{fig: bpwfs_linearity} shows both simulations, using OOMAO\cite{conan14}, and corroborating SEAL testbed results that the non-modulated bPWFS is indeed more linear than the non-modulated PWFS, both for low and high order Zernike modes.
Interestingly, one astigmatism mode shows no linearity gain in simulation and a slight degradation in lab data, which should be further investigated. The choice of modal basis also needs further investigation; initially we used TTF and then LODM actuator pokes with TTF removed (i.e., as done for LODM SHWFS control in \S\ref{sec: shwfs_wt_loop}), but simulations and lab data showed negligible bPWFS linearity enhancement with this basis. Particularly for lab data, we noticed that PWFS IM images may have had insufficient signal-to-noise ratios for the small actuator poke amplitudes needed to be within the linear regime, but where on-air turbulence averaging over longer sequences of stacked images also degraded IM quality; the optimal modal basis that can be used to acquire IMs at high enough speed to freeze on-air turbulence should there be further investigated. Alignment also served as a potential limitation in laboratory demonstrations. Although we wrote an automated alignment script designed to minimize the standard deviation of the pupil intensities over tip, tilt, and focus to find the best alignment in an automated way each time we swapped PWFS and bPWFS masks on SEAL, LO linearity curves remained offset between the two masks (hence the need to show their integrals in Fig. \ref{fig: bpwfs_linearity} for a fair comparison); more robust alignment algorithms (e.g., alignment by slope minimization) could enable a more uniform comparison between the two masks.

As discussed above, future work will need to include a more integrated plan on how to operate the non-modulated bPWFS in closed-loop within existing AO systems. Before that, future bPWFS SEAL testing goals not already mentioned include linearity analysis of higher order modes beyond the LODM Nyquist limit with our HODM and closed-loop analysis of AO residual level input WFE. For the latter, our SLM\cite{vanKooten22b} can be used for an aliasing error budget analysis, comparing DM-injected and reconstructed turbulence can be used for a non-linearities error budget analysis, and results with our tuneable light source (with a range from 0.01 to 8 mW) comparing a photon noise-limited and WFE-limited regime can be used for a photon noise error budget analysis.
\section{CONCLUSIONS AND FUTURE WORK}
\label{sec: conclusions}
In this paper we have presented laboratory results with SEAL (\S\ref{sec: setup}) on three ongoing areas of HCI and ExAO technology development:
\begin{enumerate}
    \item \S\ref{sec: fast}: Multi-WFS SCAO using a SHWFS and FAST to control a LODM and HODM. Thus far we have demonstrated separate high-speed real-time control of each WFS with both DMs but have not yet combined both WFSs into a single real-time loop. SEAL development to implement real-time temporal filtering in order to combine both WFSs is ongoing.
    \item \S\ref{sec: pupil_chopping}: Pupil chopping, both with an external optical chopper (\S\ref{sec: v1}) and internal DM-based chopping approach (\S\ref{sec: v2}). We have demonstrated the power of either approach to enable second stage AO and/or multi-WFS SCAO, in this case for a non-coronagraphic science image at a $>$50\% science duty cycle (but with coronagraphic pupil chopping investigations to come in the future).
    \item \S\ref{sec: bpwfs}: testing the bright Pyramid WFS concept. We have demonstrated that the non-modulated bPWFS is more linear than the non-modulated PWFS, but also finding that a modulated to non-modulated closed-loop bootstrapping procedure is ineffective, prompting further work needed on the feasibility of second stage and/or multi-WFS SCAO with a bPWFS.
\end{enumerate}
\acknowledgements
We gratefully acknowledge research support of the University of California Observatories and UCSC's Lamat NSF REU program for funding this research. This work also benefited from the 2022 Exoplanet Summer Program in the Other Worlds Laboratory (OWL) at the University of California, Santa Cruz, a program funded by the Heising-Simons Foundation. Author B. Gerard thanks the 2018 SCExAO team for hosting discussions that led to the pupil chopping concept and the UCSC LAO group, particularly Dominic Sanchez, for discussions that led to the DM-based pupil chopping idea. 

\bibliography{report} 

\begin{thebibliography}{10}

\bibitem{nielsen19}
{Nielsen}, E.~L., {De Rosa}, R.~J., {Macintosh}, B., {Wang}, J.~J., {Ruffio},
  J.-B., {Chiang}, E., {Marley}, M.~S., {Saumon}, D., {Savransky}, D.,
  {Ammons}, S.~M., {Bailey}, V.~P., {Barman}, T., {Blain}, C., {Bulger}, J.,
  {Burrows}, A., {Chilcote}, J., {Cotten}, T., {Czekala}, I., {Doyon}, R.,
  {Duch{\^e}ne}, G., {Esposito}, T.~M., {Fabrycky}, D., {Fitzgerald}, M.~P.,
  {Follette}, K.~B., {Fortney}, J.~J., {Gerard}, B.~L., {Goodsell}, S.~J.,
  {Graham}, J.~R., {Greenbaum}, A.~Z., {Hibon}, P., {Hinkley}, S., {Hirsch},
  L.~A., {Hom}, J., {Hung}, L.-W., {Dawson}, R.~I., {Ingraham}, P., {Kalas},
  P., {Konopacky}, Q., {Larkin}, J.~E., {Lee}, E.~J., {Lin}, J.~W., {Maire},
  J., {Marchis}, F., {Marois}, C., {Metchev}, S., {Millar-Blanchaer}, M.~A.,
  {Morzinski}, K.~M., {Oppenheimer}, R., {Palmer}, D., {Patience}, J.,
  {Perrin}, M., {Poyneer}, L., {Pueyo}, L., {Rafikov}, R.~R., {Rajan}, A.,
  {Rameau}, J., {Rantakyr{\"o}}, F.~T., {Ren}, B., {Schneider}, A.~C.,
  {Sivaramakrishnan}, A., {Song}, I., {Soummer}, R., {Tallis}, M., {Thomas},
  S., {Ward-Duong}, K., and {Wolff}, S., ``{The Gemini Planet Imager Exoplanet
  Survey: Giant Planet and Brown Dwarf Demographics from 10 to 100 au},'' {\em
  Astronomical Journal}~{\bf 158},  13 (July 2019).

\bibitem{jensen-clem22}
Jensen-Clem, R., Hinz, P., Skemer, A., Wizinowich, P., Jovanovic, N., Mazin,
  B.~A., John I.~Bailey, I., Frazin, R.~A., Sallum, S., Males, J.~R., and
  Tamura, M., ``{A Technology and Science Gap List for Habitable-Zone Exoplanet
  Imaging with Ground-Based Extremely Large Telescopes},'' in [{\em In this
  proceeding}{\nolinebreak\hspace{0.1em}]},  {\em Society of Photo-Optical
  Instrumentation Engineers (SPIE) Conference Series} (July 2022).

\bibitem{jensen-clem21}
{Jensen-Clem}, R., {Dillon}, D., {Gerard}, B., {van Kooten}, M.~A.~M.,
  {Fowler}, J., {Kupke}, R., {Cetre}, S., {Sanchez}, D., {Hinz}, P., {Laguna},
  C., {Doelman}, D., and {Snik}, F., ``{The Santa Cruz Extreme AO Lab (SEAL):
  design and first light},'' in [{\em Society of Photo-Optical Instrumentation
  Engineers (SPIE) Conference Series}{\nolinebreak\hspace{0.1em}]},  {\em
  Society of Photo-Optical Instrumentation Engineers (SPIE) Conference Series}
  {\bf 11823},  118231D (Sept. 2021).

\bibitem{gerard22}
Gerard, B.~L., Dillon, D., Cetre, S., and Jensen-Clem, R.~M., ``{Laboratory
  demonstration of real-time focal plane wavefront control of residual
  atmospheric speckles},'' {\em Journal of Astronomical Telescopes,
  Instruments, and Systems}~{\bf 8}(3),  1 -- 22 (2022).

\bibitem{sanchez22}
Sanchez, D.~F., Hinz, P., and Dillon, D., ``{Preliminary lab demonstration of a
  3-sided reflective pyramid wavefront sensor for Shane AO using SEAL
  testbed},'' in [{\em In this proceeding}{\nolinebreak\hspace{0.1em}]},  {\em
  Society of Photo-Optical Instrumentation Engineers (SPIE) Conference Series}
  (July 2022).

\bibitem{salama22}
Salama, M., Jensen-Clem, R., van Kooten, M., Dillon, D., Gerard, B.~L., Fowler,
  J., Cetre, S., Snik, F., and Doelman, D., ``{Vector Zernike Wavefront Sensor
  on the Santa Cruz Extreme AO Lab (SEAL) Testbed},'' in [{\em In this
  proceeding}{\nolinebreak\hspace{0.1em}]},  {\em Society of Photo-Optical
  Instrumentation Engineers (SPIE) Conference Series} (July 2022).

\bibitem{vanKooten22b}
van Kooten, M.~A., Jensen-Clem, R., Fowler, J., Dillon, D., Kupke, R., Salama,
  M., and Gerard, B.~L., ``{Spatial light modulator on Santa Cruz Extreme AO
  Laboratory (SEAL) testbed},'' in [{\em In this
  proceeding}{\nolinebreak\hspace{0.1em}]},  {\em Society of Photo-Optical
  Instrumentation Engineers (SPIE) Conference Series} (July 2022).

\bibitem{cetre18}
{Cetre}, S., {Guyon}, O., {Bond}, C., {Chun}, M., {Mawet}, D., {Wizinowich},
  P., {Lockhart}, C., {Goebel}, S., and {Wetherell}, E., ``{A near-infrared
  pyramid wavefront sensor for Keck adaptive optics: real-time controller},''
  in [{\em Adaptive Optics Systems VI}{\nolinebreak\hspace{0.1em}]},  {Close},
  L.~M., {Schreiber}, L., and {Schmidt}, D., eds., {\em Society of
  Photo-Optical Instrumentation Engineers (SPIE) Conference Series} {\bf
  10703},  1070339 (July 2018).

\bibitem{sengupta22}
Sengupta, A.~R., Gerard, B.~L., Dillon, D., van Kooten, M., Gavel, D., and
  Jensen-Clem, R., ``{Laboratory Demonstration of Optimal Identification and
  Control of Tip-Tilt Systems},'' in [{\em In this
  proceeding}{\nolinebreak\hspace{0.1em}]},  {\em Society of Photo-Optical
  Instrumentation Engineers (SPIE) Conference Series} (July 2022).

\bibitem{gerard21a}
Gerard, B.~L., Véran, J.-P., Singh, G., Herriot, G., Lardière, O., and
  Marois, C., ``{Fast focal plane wavefront sensing as a second stage adaptive
  optics wavefront sensor},'' in [{\em Adaptive Optics Systems
  VII}{\nolinebreak\hspace{0.1em}]},  Schreiber, L., Schmidt, D., and Vernet,
  E., eds.,  {\bf 11448},  483 -- 499, International Society for Optics and
  Photonics, SPIE (2021).

\bibitem{gerard21b}
Gerard, B.~L., Chambouleyron, V., Jensen-Clem, R., and Sauvage, J.-F., ``{The
  bright pyramid wavefront sensor},'' in [{\em Techniques and Instrumentation
  for Detection of Exoplanets X}{\nolinebreak\hspace{0.1em}]},  Shaklan, S.~B.
  and Ruane, G.~J., eds.,  {\bf 11823},  403 -- 412, International Society for
  Optics and Photonics, SPIE (2021).

\bibitem{gavel14}
{Gavel}, D. and {Norton}, A., ``{Woofer-tweeter deformable mirror control for
  closed-loop adaptive optics: theory and practice},'' in [{\em Adaptive Optics
  Systems IV}{\nolinebreak\hspace{0.1em}]},  {Marchetti}, E., {Close}, L.~M.,
  and {Vran}, J.-P., eds., {\em Society of Photo-Optical Instrumentation
  Engineers (SPIE) Conference Series} {\bf 9148},  91484J (Aug. 2014).

\bibitem{vanKooten22a}
{van Kooten}, M. A.~M., {Jensen-Clem}, R., {Cetre}, S., {Ragland}, S., {Bond},
  C.~Z., {Fowler}, J., and {Wizinowich}, P., ``{Predictive wavefront control on
  Keck II adaptive optics bench: on-sky coronagraphic results},'' {\em Journal
  of Astronomical Telescopes, Instruments, and Systems}~{\bf 8},  029006 (Apr.
  2022).

\bibitem{guyon17}
{Guyon}, O. and {Males}, J., ``{Adaptive Optics Predictive Control with
  Empirical Orthogonal Functions (EOFs)},'' {\em arXiv e-prints} ,
  arXiv:1707.00570 (July 2017).

\bibitem{martinache13}
{Martinache}, F., ``{The Asymmetric Pupil Fourier Wavefront Sensor},'' {\em
  Publications of the Astronomical Society of the Pacific}~{\bf 125},  422
  (Apr. 2013).

\bibitem{johnson22}
Johnson, A.~B., Marois, C., Gamroth, D., Lardi\`{e}re, O., Thompson, W., Singh,
  G., Fitzsimmons, J., and Bradley, C., ``{Blinking the fringes, initial
  development and results of the Ultra-Low Speed Optical Chopper for the
  Self-Coherent Camera},'' in [{\em In this
  proceeding}{\nolinebreak\hspace{0.1em}]},  {\em Society of Photo-Optical
  Instrumentation Engineers (SPIE) Conference Series} (July 2022).

\bibitem{codona13}
{Codona}, J.~L., ``{Differential OTF Wavefront Sensing},'' {\em arXiv e-prints}
  ,  arXiv:1309.6719 (Sept. 2013).

\bibitem{fu21}
Fu, Q., Amata, H., Gerard, B., Marois, C., and Heidrich, W., ``{Additive
  lithographic fabrication of a Tilt-Gaussian-Vortex mask for focal plane
  wavefront sensing},'' in [{\em Optifab 2021}{\nolinebreak\hspace{0.1em}]},
  Nelson, J.~D. and Unger, B.~L., eds.,  {\bf 11889},  162 -- 170,
  International Society for Optics and Photonics, SPIE (2021).

\bibitem{conan14}
{Conan}, R. and {Correia}, C., ``{Object-oriented Matlab adaptive optics
  toolbox},'' in [{\em Adaptive Optics Systems
  IV}{\nolinebreak\hspace{0.1em}]},  {Marchetti}, E., {Close}, L.~M., and
  {Vran}, J.-P., eds., {\em Society of Photo-Optical Instrumentation Engineers
  (SPIE) Conference Series} {\bf 9148},  91486C (Aug. 2014).

\end{thebibliography}
\bibliographystyle{spiebib} 

\end{document}